\documentclass{iau} 
\usepackage{graphicx}
\usepackage{wrapfig}
\usepackage[colorlinks=true,citecolor=black]{hyperref}

\title
{Stellar Winds in Massive X-ray Binaries}

\author[P.~Kretschmar, S.~Mart\'inez-N\'u\~nez, E.~Bozzo, et al.]   
{Peter Kretschmar$^1$, 
 Silvia Mart\'inez-N\'u\~nez$^2$, 
 Enrico Bozzo$^3$, 
 Lidia M.\ Oskinova$^4$, 
        Joachim Puls$^5$, 
        Lara Sidoli$^6$, 
        Jon Olof Sundqvist$^7$, 
        Pere Blay$^8$, 
        Maurizio Falanga$^9$, 
        Felix F\"urst$^1$, 
        Angel G\'{\i}menez-Garc\'{\i}a$^{10}$, 
        Ingo Kreykenbohm$^{11}$, 
        Matthias K\"{u}hnel$^{11}$, 
        Andreas Sander$^4$, 
        Jos\'e Miguel Torrej\'on$^{10}$, 
        J\"orn Wilms$^{11}$, 
        Philipp Podsiadlowski$^{12}$ \and
        Antonios Manousakis$^{13,14}$}

\affiliation{%
$^1$European Space Astronomy Centre (ESA/ESAC), Science Operations Department\\
P.O. Box 78, E-28691, Villanueva de la Ca\~{n}ada, Madrid, Spain
\\ email: \texttt{peter.kretschmar@esa.int} \\[\affilskip]
$^2$Instituto de F\'isica de Cantabria (CSIC-Universidad de Cantabria)
E-39005, Santander, Spain\\[\affilskip]
$^3$ISDC, University of Geneva, Chemin d’Ecogia 16, Versoix, 1290, Switzerland\\[\affilskip]
$^4$Institut f\"ur Physik und Astronomie, Universit\"at Potsdam,\\
Karl-Liebknecht-Str. 24/25, D-14476 Potsdam, Germany\\[\affilskip]
$^5$Universit\"atssternwarte der Ludwig-Maximilians-Universit\"at M\"unchen,\\
Scheinerstrasse 1, 81679, M\"unchen, Germany\\[\affilskip]
$^6$INAF, Istituto di Astrofisica Spaziale e Fisica Cosmica - Milano,\\ via E. Bassini 15, I-20133 Milano, Italy \\[\affilskip]
$^7$Instituut voor Sterrenkunde, KU Leuven, Celestijnenlaan 200D, 3001 Leuven, Belgium\\[\affilskip]
$^8$Nordic Optical Telescope - IAC, P.O.Box 474, E-38700, Santa Cruz de La Palma\\
Santa Cruz de Tenerife, Spain\\[\affilskip]
$^9$International Space Science Institute (ISSI), Hallerstrasse 6, CH-3012 Bern, Switzerland \\[\affilskip]
$^{10}$Instituto Universitario de F\'isica Aplicada a las Ciencias y las Tecnolog\'ias,\\
University of Alicante, P.O. Box 99, E03080 Alicante, Spain \\[\affilskip]
$^{11}$ Dr. Karl Remeis-Observatory \& ECAP, Universit{\"a}t Erlangen-N\"urnberg, \\Sternwartstr. 7, D-96049 Bamberg, Germany \\[\affilskip]
$^{12}$Department of Astronomy, Oxford University, Oxford OX1 3RH, United Kingdom\\[\affilskip]
$^{13}$Centrum Astronomiczne im. M. Kopernika, Bartycka 18, 00-716, Warszawa, Poland\\[\affilskip]
$^{14}$Department of Physics, Sultan Qaboos University, 123 Muscat, Oman  
}

\pubyear{2017}
\volume{329}  
\jname{The lives and death-throes of massive stars}
\editors{A.C. Editor, B.D. Editor \& C.E. Editor, eds.}
\begin{document}

\maketitle

\begin{abstract}
Strong winds from massive stars are a topic of interest to a wide range of astrophysical fields. In High-Mass X-ray Binaries the presence of an accreting compact object on the one side allows to infer wind parameters from studies of the varying properties of the emitted X-rays; but on the other side the accretor's gravity and ionizing radiation can strongly influence the wind flow. Based on a collaborative effort of astronomers both from the stellar wind and the X-ray community, this presentation attempts to review our current state of knowledge and indicate avenues for future progress.
\end{abstract}

\firstsection 
\section{Structures in winds from massive stars}
Winds from massive stars are attributed to radiative line-driving, see, e.g., \cite{Puls08} for a review.
Although the standard theory of line-driven winds assumes a stable, time-independent and homogeneous wind,
both theoretical considerations and observational features at different wavelengths clearly indicate 
that the winds of massive stars are not smooth and isotropic, but structured.

\textit{Small-scale structures} are explained by reverse shocks in the wind, which are
caused by a very strong, intrinsic instability in line-drive winds (LDI), already noted by
\cite{LS70}. Numerical hydrodynamical modelling, e.g., by \cite{Feldmeier95} or \cite{Sundqvist13}
finds that the wind plasma becomes compressed into spatially narrow ``clumps''
separated by large regions of rarefied gas. 
The characteristic length scale for these 
structures is the Sobolev length; for typical hot supergiants this leads to an order of magnitude estimate 
of 10$^{18}$\,g for typical clump masses and a few $R_{\odot}$ for their extent.
See, e.g., \cite{OskFeldKre:2012} for specific predictions.

\textit{Large-scale structures} in winds from massive stars are mainly inferred from the 
so-called Discrete Absorption Components (DACs), observed in most O- and early B-star winds 
(\cite[Howarth \& Prinja 1989]{HP89}) and in late B-supergiants 
(\cite[Bates \& Gilheany 1990]{BatesGilheany90}). A 
widely held candidate mechanism for these structures are Co-rotating Interaction Regions
(\cite[Mullan 1984,1986]{Mullan84,Mullan86}), well studied in the solar wind. Another candidate
are Rotational Modulations (RMs), as reported, e.g., by \cite{Massaetal95}. The density
contrasts for these larger structures are rather low (factors of at most a few), but they may contain 
large overall masses, e.g., $10^{21-22}$\,g for similar assumptions as above.

\section{Wind-accreting High-Mass X-ray Binaries}
In High-Mass X-ray Binaries (HMXB) a compact object, mostly a neutron star, sometimes a black hole
or a white dwarf accretes matter from its companion and produces copious X-ray radiation. For a
typical neutron star $L_X \approx 0.1 \dot{M}c^2$ for a mass accretion rate $\dot{M}$. There are several mechanisms 
to fuel the X-ray source, e.g., Roche-Lobe overflow, or from the disk around a Be star
-- neither discussed further here -- but also accretion from the massive star wind. This last mechanism
is present in two sub-groups: Classical Supergiant X-ray Binaries (SGXB) tend to be mostly persistent 
sources with erratic variations in flux. The more recently identified sub-group of Supergiant Fast 
X-ray Transients (SFXTs) has similar system parameters (where known), but remains mostly in a low luminosity state 
with brief outbursts and much larger flux variations. For a recent overview of different HXMB in our Galaxy 
see \cite{Walter:2015}.

\section{X-ray absorption and fluorescence}
A conceptually straightforward method to infer clumps or larger structures in stellar winds is to measure 
the attenuation of the X-ray flux, i.e., the variations in the measured absorbing column which in HMXB usually 
is in the range $N_\mathrm{H} \sim 10^{21-24}$ cm$^{-2}$.
The main caveat is that this requires a good knowledge of the unabsorbed spectral continuum 
in order to minimise the degeneracy between spectral slope and absorption. An implicit issue is also that 
accreting X-ray sources are intrinsically variable and thus care has to be taken when comparing different 
observations. 
To obtain detailed observational results on wind structures, very extensive campaigns are required, like that 
reported in \cite{grinberg2015} and previous publications for Cyg X-1. Large scale structures can also
be traced in some cases with the lower time resolution of X-ray monitor data as, e.g., \cite{Malacaria:2016}
have demonstrated. 

Another diagnostic is from X-ray fluorescence lines which will stem mostly from emission nearby to
the compact object at most a few $R_\odot$ from the X-ray source. The line parameters
can yield information on distribution, velocities and ionisation of the reprocessing material
as described, e.g., in \cite{Gimenez2015} and references therein.

\section{Tracing accreted mass}

As explained above, the X-ray luminosity of an accreting compact object is a direct measure of the
current mass accretion rate. Assuming direct infall of matter, the X-ray source would then be
a ``local probe'' of structures in the wind traversed by it. This approach has been used by various
authors to explain flares and low-flux or ``off'' states in HMXB, e.g., by \cite{Ducci2009} or
\cite{Fuerst:2010}. But the estimates for clump masses from such studies have sometimes been 2--3
order of magnitudes larger than those from hydrodynamical simulations of stellar winds.

A closer look at accretion physics also shows that direct infall of captured matter is not
necessarily taking place. According to \cite{OskFeldKre:2012}, this would also imply
orders of magnitude higher variability in many systems than observed.
Different studies in recent years discuss, e.g., the possibility of
settling envelopes
around the compact objects, depending on conditions (\cite[Shakura \etal\ 2012]{Shakura2012}).
Another possibility is Chaotic Cold Accretion with complex accretion flows and condensation
to filaments and cool clumps, as \cite{Gaspari:2017} have put forward as a model for AGNs, 
noting that the findings may also apply to X-ray binaries. In the common case of an accreting 
neutron star with a strong magnetic field, the interaction with the magnetosphere will lead to
additional complications, including possible inhibition of accretion as detailed, e.g.,
in \cite{Bozzo2008}.

\section{Feedback on wind flow}
The presence of the X-ray emitting compact object evidently also influences the wind flow, sometimes quite
dramatically so. The gravitational pull focusses the stellar wind in the orbital plane. The bow shock
of the compact object moving through the dense wind can create an ``accretion wake'' following the
compact object in its orbit. Also, the intense X-ray
emission of bright sources creates a large Str{\"o}mgren sphere in which the wind is photoionised and the
wind acceleration can be slowed or even cut off. These effects 
have been discussed in quite some
detail already by Blondin \etal\ (e.g., \cite[1990]{Blondin:90}); for recent simulations of these effects see, e.g.,
\cite{ManousakisWalter:2015a} or \cite{Cechura:2015}. 
But so far, these feedback models have been based on smooth winds, while
models including LDI and clumpy winds have usually not included an accretor and X-ray feedback.


\section{Ongoing efforts and Outlook}
The authors of this contribution and other colleagues have met at the International Space Science Institute (ISSI) Bern for meetings in \href{http://www.issibern.ch/teams/stellarwind/}{2013 \& 2014} and a differently
structured follow-up group is meeting again in \href{http://www.issibern.ch/teams/stellarwindxray/}{2016 \& 2017}
in order to discuss the open questions and possible avenues forward. 

Among the findings of the first series of meetings are: (1) serious discrepancies in clump sizes and density
contrasts used in the literature; (2) systematically lower wind velocities (factor 2--5) in HMXB than those 
derived for single stars; (3) CIRs should be stable over several orbits, but this is not reflected in HMXB 
studies of orbital variation; (4) the different behaviour of classical SGXBs and SFXTs remains an open question
with no simple explanation. These findings and other results have been published in a detailed review by 
\cite{ISSIReview}. The ongoing meetings aim to reduce some of the uncertainties recognised in the
first set, as well as include more modelling efforts for wind structure and accretion, and also discuss the impact
of these findings for population synthesis studies.

\begin{wrapfigure}[14]{r}{0.64\textwidth}
\vspace*{-3mm}
\includegraphics[angle=0,width=0.64\textwidth]{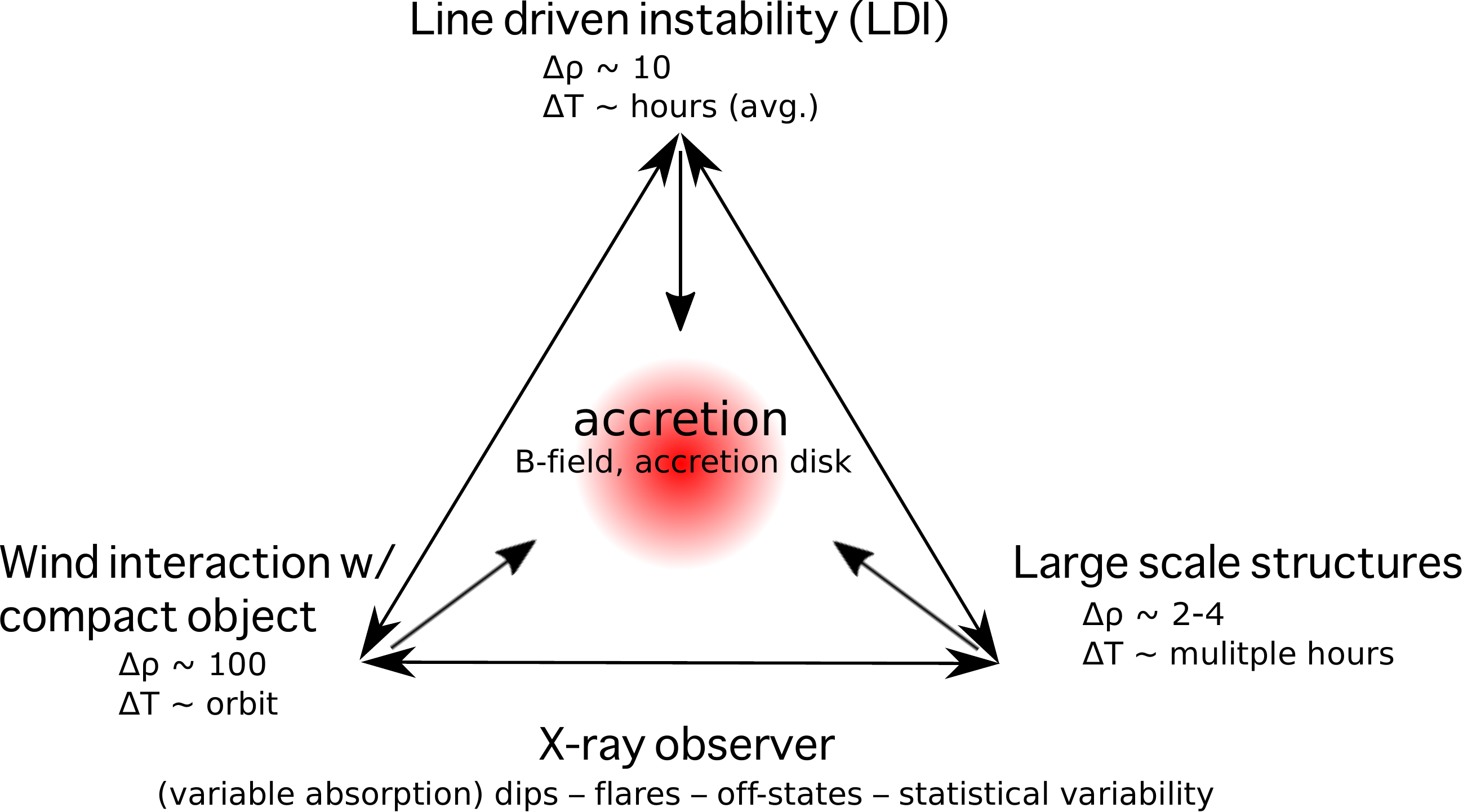}\vspace*{-1.5mm}
\caption{Scheme of interactions in a HMXB, $\Delta\rho$ indicates typical density variations and $\Delta$T typical time scales.}
\label{fig:interact}
\end{wrapfigure}
For the future, we hope to arrive at models com\-bi\-ning intrinsically clumpy winds with the effects 
from X-ray feedback, including a realistic picture of time 
varying accretion and X-ray emission. Sys\-te\-matic
multi-wavelength observations via coordinated campaigns with space and ground instruments are
required to follow variations on time scales of days or faster. The arrival of fast, sensitive optical
spectrographs on ground allows to study some wind variations on time scales of seconds. In space, the
advent of X-ray calorimeters will open a new era of X-ray line diagnostics. Until that time, further deep, 
dedicated observations with the existing grating instruments could still shed light on many questions.

\end{document}